# Pairing gap as a new observable for critical points in the region of A≈100

Hossein Emami[a], Hadi Sabri[b]

Faculty of Physics, University of Tabriz, Tabriz, Iran

[a] e-mail: hosseinemami930@yahoo.com (Corresponding author)
[b] e-mail: h-sabri@tabrizu.ac.ir

## Abstract

This study used the pairing gap to identify nuclei as candidates for critical point symmetry around Z≈40 and A≈100. Nuclei around A ≈ 100 display complex shape evolution and configuration crossing patterns. We utilized the experimental and algebraic frameworks of the interacting boson model and the newly developed interacting boson-fermion model to study the isotopes of Mo and Ru. The results show significant variations in these quantities across nuclei located at the E(5) and X(5) critical points, analyzed through different observables. We also examined another region that is suitable for a shape phase transition. Furthermore, our findings suggest new candidates for critical points in other phase transitional regions for different isotopic chains.

Keywords: quantum phase transition, critical point symmetry, pairing gap, d-pairing gap, neutron number.

## 1. Introduction:

Quantum Phase Transition (QPT) and Critical Point Symmetries (CPS) have some microscopic and macroscopic (observable) signatures, and both signatures are related to the shape and structure of nuclei. Nuclei around Z ≈ 40 and A ≈ 100 have been known for a long time to show a sudden change from spherical to deformed ground states, and a sudden change in the properties of the ground state is the experimental signal for the shape phase transition (SPT) in nuclei [1,2]. QPT and CPS suggest a sudden alteration in the characteristics of the ground state to change abruptly, leading to the quick change of various observables, such as two neutron separation energies($S_{2n}$)[3-13], $E(4_1^+)/E(2_1^+)$[3-4,13-15], $B(E2)$[3,5,11,13,14,16,17], isotopic shifts[3-5,13], β-γ order parameters[3,5,13-17], neutron capture cross section[7], hindrance factors [18], and from some other signatures can be mentioned: Level density parameter[4], Life time[19,20], spectroscopic quadrupole moments[21]. In transitional nuclei near the CPS, X(3) and Z(4), there are two quantum concepts called the minimal length and the deformation-dependent mass. A correlation like this could be regarded as a new signature for these CPS, which enabled us to predict new candidate nuclei for these critical points [22]. Therefore, it is possible to study different structures of nuclei with all the signatures mentioned above. Atomic nuclei can be deformed, and this deformation can be attributed to the features of nuclear structures, including compounds that can affect the deformation, pairing gap effects, and quadrupole-quadrupole (Q-Q) interaction. The most important interactions for short-range correlations are pairing interactions, whereas for long-range correlations, Q-Q interactions hold greater significance. [23]. By comparing the pairing gap to the Q-Q interaction, we observe a strong correlation in pairing that leads to competition. This competition can be considered in the context of quantum phase transitions (QPT). As the pairing gap increases, it tends to approach a spherical state. On the other hand, increasing the Q-Q interaction leads to a deformed state. Therefore, whichever parameter (pairing gap or Q-Q interaction) is more robust, it alters the shape of the nucleus in its favor. This observation led us to the idea that the pairing gap can be a signature of critical points and QPT in different isotopic chains.

The nuclear shell structure is enhanced by pair correlations [24,25]. The Bardeen, Cooper, and Schrieffer (BCS) [26] approximation is often used to treat pairing [27-29]. Pairing correlations are characterized by an energy gap in the excitation spectrum [30]. The pairing gap in atomic nuclei is around 1-2 MeV, compared to the typical energy scale of the N-N interaction, which is a few hundred MeV [31]. Different formulas were used to select pairing gaps between nucleons, including those obtained from even-odd mass differences reflected in the liquid-drop term described by Bohr and Mottelson [32]:

$$\Delta_{BM} \approx 12A^{-1/2}, \tag{1}$$

Equation (1) yields the same value for neutron-neutron and proton-proton pairings [33]. Other formulas for calculating the pairing gaps are based on binding energy (BE) and separation energies, as given by the following:

$$\Delta(Z, N) = \frac{-1^N}{2} [2BE(Z,N) - BE(Z,N-1) - BE(Z,N+1)], \tag{2}$$

$$\Delta(Z, N) = \frac{-1^{N+1}}{2} [S_n(Z,N+1) - S_n(Z,N)], \tag{3}$$

Estimating the pairing gap based on the spectral properties of a nucleus is a common approach, but certain definitions of the pairing gap cannot be applied to closed-shell nuclei. Consequently, it is safest to calculate the pairing gaps from mean-field calculations using the same method as the experimental values [30].

The pairing gaps play a significant role in the proton-neutron quasiparticle random phase approximation (pn-QRPA), and Eq.(2) led to the most accurate prediction of β-decay half-lives [33], and a study about the thermodynamic features of pairing within many-body systems[34]. Pairing gaps in mean-field configurations impact NMEs of 0νββ decay. [35], The study about the partial-wave nuclear force contributes to pairing in nuclei at the level of pairing matrix elements [36] and alpha decay properties [37]. In some studies, the pairing gap has been partially used to describe the SPT in the nucleus [35].

QPT differs from thermodynamic transitions in that it involves the equilibrium shape changes in the ground state of nuclei at absolute zero temperature. SPT or ground state phase transition is also used to refer to it, although it can also be applied to excited states [38-40]. Most experimental and theoretical studies on first and second-order nuclear QPTs have examined systems with even numbers of protons and neutrons. Shape coexistence (SC) refers to the specific situation in which the ground-state band of the nucleus is close to another band with a completely different structure. In even-even nuclei, shape coexistence often leads to the presence of a $0^+$ band that is closely situated in energy to the ground state band, yet possesses a fundamentally different structure. For instance, one of the bands may be spherical while the other is deformed. Thus, the nature of low-lying $0^+$ bands in even-even nuclei is of interest [41]. Most studies have focused on experimental observations and relevant theoretical developments [41-46]. QPT and SC can be related to each other, and some investigations about the connection between QPT and SC have been given in Refs. [3,41,47-51]. Important signatures and the effect of SC are Strong electric monopole transitions, characterized by the monopole strength, connecting excited $0^+$ states to the ground state [52], determining the half−lives[53]. Some other signatures for SC and SPT have been given in Ref. [47]. The study of transitions from one phase to another was facilitated by this fact, which led to the creation of CPS for these phase transitions [54]. In nuclear physics, there are two CPS known as the E(5) [55] and X(5) symmetries [56]. The E(5) symmetry is believed to correspond to the transition from vibrational U(5) to γ-unstable O(6) [57] nuclei, while the X(5) symmetry is assumed to describe the transition from vibrational U(5) to prolate axially symmetric SU(3) nuclei [22,58]. Notable solutions of the Bohr Hamiltonian yield both symmetries [59]. CPS [58] describes nuclei at the points of SPT between different limiting symmetries; recent attention has been directed towards them because they produce parameter-independent predictions and are in good agreement with experiments [60-64].

Eq. (2) is based on the BE, and the odd-even oscillation in BE as a function of neutron number is one of the most robust signatures of pairing in nuclei [65]. The BE can be expressed as an analytical function based on the number of nucleons [30]. Accurate estimation of the shell correction energy

[66] is essential for the precise determination of BE, level density, and other structural properties of nuclear systems [67]. Observables such as nuclear masses and BE can be used to characterize a nucleus and obtain information about nuclear correlations [9]. The nuclear masses can be used to compute several quantities that are crucial for understanding the various aspects of nuclear structure. In particular, the quantity of interest is $S_{2n}$ from even-even medium mass nuclei, which subtracts BE and provides some information about the nuclear structure [10]. $S_{2n}$ is an observable that depicts the first and second order of QPTs [11], $S_{2n}$ is a direct and primary signature of the emergence of the SPT [68]. Various studies and calculations add parameters about ($S_{2n}$) or ($dS_{2n}$) to understand different aspects of nuclear structure, as discussed in Refs. [3-12,69,70]. According to the relation between $S_{2n}$ and BE, and also according to the pairing gap and its relation with BE, we can suggest similarly that the pairing gap can be related to $S_{2n}$. Similarly, according to the empirical correlation between neutron capture cross-sections and $S_{2n}$ [71,72] and also knowing that the neutron capture cross-section is one of the several observables of QPT [6], the neutron capture cross-section can be related to the pairing gap in the nucleus. In the following, we also use some advanced theoretical formulas about the binding energy and pairing gap in some algebraic models such as the Interacting Boson Model (IBM) and Interacting Boson-Fermion Model (IBFM), to show, how changes the pairing gap quantities, and compare these quantities with experimental values of the pairing gap.

## 2. Results and Theoretical Framework

Our focus in this article is on the evolution of the pairing gap as a function of neutron number, by using empirical data (taken from [73]), and we will compare it to the theoretical calculation in the IBM and IBFM frameworks. In the following, we will suggest that the evolution of the pairing gap can be a signature of critical points in different isotopic chains, indicating that the evolution of the pairing gap can be a signature of critical points in the region of A≈100. We have used their differential variation d(pairing gap), instead of the pairing gap, because the d(pairing gap) has a straightforward dependence and is more sensitive to neutron numbers as an important control parameter for the phase transition.

In this article, we are investigating Mo and Ru isotopic chains that are near in the region of A≈100, which is suitable for studying QPT and CPS [13,74-76]. Nuclei that have a mass number around 100 and an atomic number close to 40 are thought to undergo an abrupt transition in the arrangement of their ground state and non-yrast state as the number of neutrons varies [75,76]. Investigating even-even and neighboring odd-mass nuclei enhances our understanding of the development of deformation and shape-phase transitions. CPS reveals that some regions of the nuclear chart display rapid transitions between symmetries [58]. Various studies about QPT and CPS reported the best regions for a study about E(5) and X(5) symmetry, so that these regions are in the near Z ≈ 40-82 [3,5,6,11,16,38,47,56]. Similarly, in the following, we will extend our idea about the pairing gap as a new observable to other isotopic chains that are located in these regions, such as the Nd, Hf, Os, and Pt isotopic chains. Since the QPT generally appears more in even nuclei, we have used different even-even isotopic chains only in this article.

## 2.1. Results of the variation of the pairing gap, d(pairing gap), and ($R_{4_1^+/2_1^+}$ versus the pairing gap) in relation to neutron number

The pairing gap is an energy gap of about 1 MeV, which exists between the ground state and nearly degenerate states with spin and parity ($J^\pi$) values of $0^+$ and $2^+, 4^+, 6^+$, and so on. [23]. In some nuclei, nucleons are found in pairs of opposite spin and angular momentum, a configuration known as a pairing state. This pairing state is energetically favorable because it lowers the overall energy of the nucleus. The pairing gap comes from the fact that it takes energy to break these pairs of nucleons. The interaction between nucleons leads to pairing gaps. Nuclei with a larger pairing gap are more stable and have longer lifetimes, while nuclei with smaller pairing gaps are more likely to undergo nuclear reactions. Pairing is crucial in exotic nuclei and weakly bound nuclei with a chemical potential magnitude comparable to that of the pairing gap [77].

According to the $dS_{2n}$ calculated in Ref. [9,12,69], and the relation between $S_{2n}$ and BE, we can similarly define the d(pairing gap): $d\Delta$ from Eq.(2). Also according to the definitions for $S_{2n}$ and $dS_{2n}$ in [4], similar to that, we note similar definitions for $\Delta(Z,N)$ and $d\Delta(Z,N)$. Thus, the pairing gap, like $S_{2n}$, plays a role in the description of the QPT. So we can define the d(pairing gap) as:

$$d\Delta(Z,N) \equiv \Delta(Z,N+1) - \Delta(Z,N), \qquad (4)$$

The pairing gap and d(pairing gap) are very sensitive parameters to nuclear structure. Thus, according to this sensitivity, the pairing gap occurs in different transitional regions. In this context, we focus on which region is important for the study of phase transition.

In this paper, we utilize the pairing gap and its derivative (Eq. 4) as observables related to nuclear structure. The variation of these quantities in Mo and Ru isotopic chains is presented in Figures 1&2, respectively. Although the variation of both observables yields the same results for CPS (shown with blue dots), Similar to Refs. [9,12] that indicated the sensitivity of $dS_{2n}$ and according to the relation between the pairing gap and $S_{2n}$ due to the BE, we defined the d(pairing gap) to show the critical point of the transition from spherical to deformed shapes, because the pairing gap quantity quite associated to each of nuclei but d(pairing-gap) is due to the difference between two nuclei, thus d(pairing gap) is more sensitive to neutron numbers.

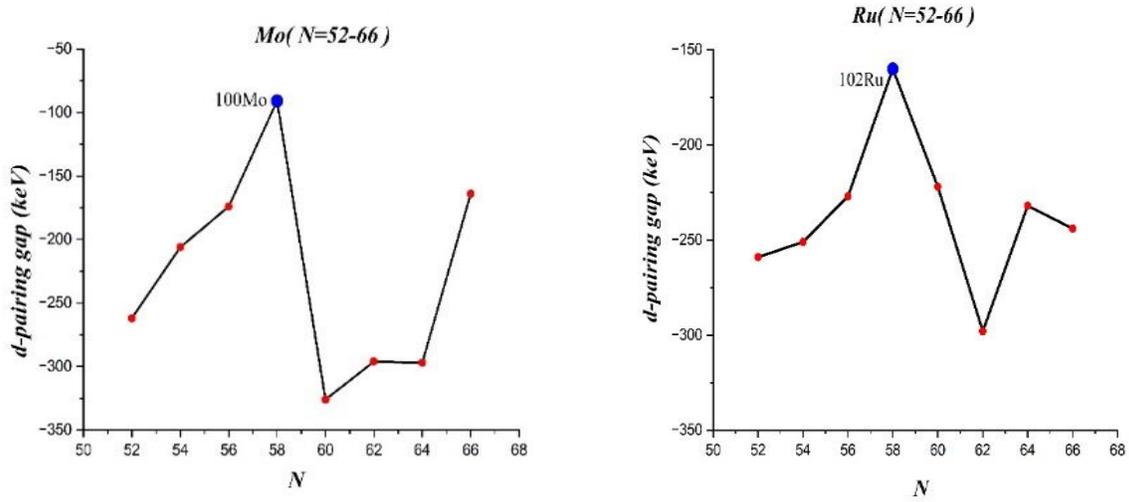

Figure 1. The variation of d(pairing gap) versus neutron number in Mo and Ru isotopic chains. Results of different studies that reported the possibility of critical points, shown with blue dots (taken from [73]).

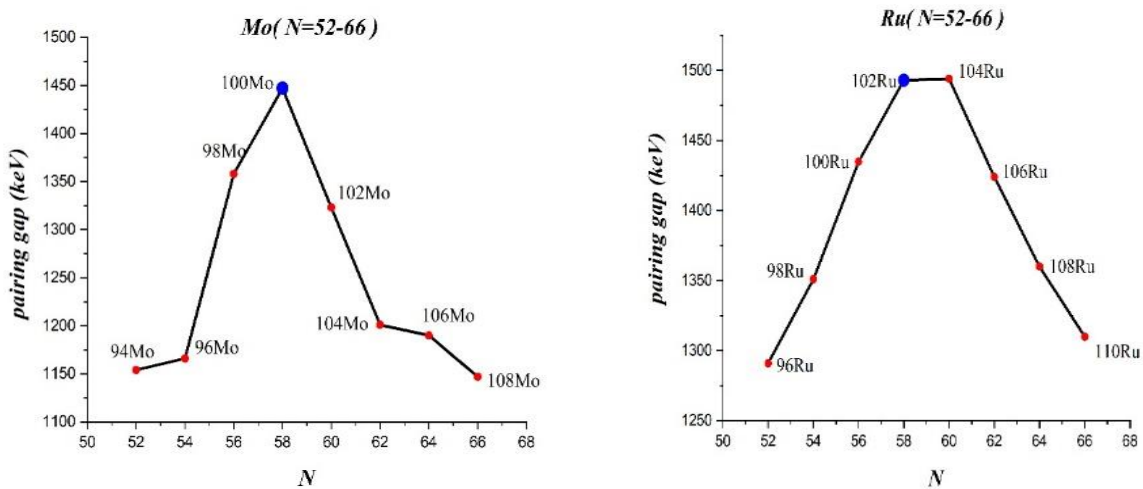

Figure 2. The variation of the pairing gap versus neutron number in Mo and Ru isotopic chains. The results of different studies that reported the possibility of critical points are shown with blue dots (taken from [73]).

The nuclear structure was described with more symmetries, such as the X(5) and E(5), which are called CPS. The energy ratio $R_{4_1^+/2_1^+} = \frac{E(4_1^+)}{E(2_1^+)}$ of the first two excited states of the ground state band is also shown, since it is a well-known and easily measurable indicator of collectivity, with deformed nuclei having $R_{4/2} > 3$, transitional nuclei exhibiting $2.4 < R_{4/2} < 3$, and vibrational nuclei possessing $R_{4/2} < 2.4$, that the $R_{4/2}$ ratio, which is 2.9 for X(5) and 2.2 for E(5) CPS [47,78]. According to importance of the region A ~ 100 for studying QPT [1-2,13,74-76,79-80], the ratio $R_{4/2}$ versus neutron number using contour plot method in term of the pairing gap for various elements in the A ~ 100 region is illustrated in Fig. 3. According to Ref. [74] and the crossing pattern [79], $R_{4/2}$ values for Z < 44 shift from being below those with Z ≥ 44 to exceeding them between N = 58 and 60.

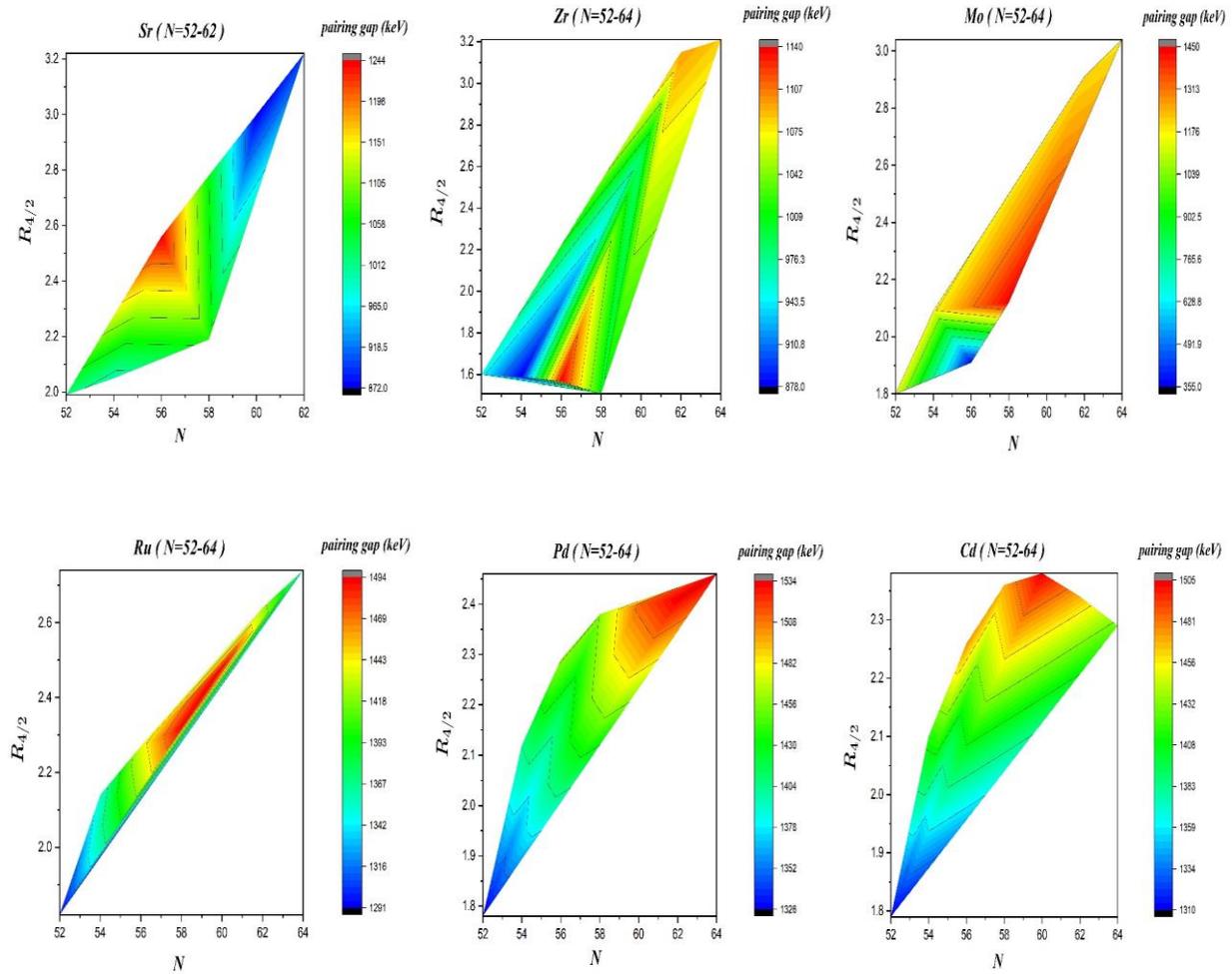

Figure 3. Contour plots in the ($R_{4/2} = \frac{E(4_1^+)}{E(2_1^+)}$, neutron number) plane, in terms of the pairing gap, the near A~100 region (data taken from [73]).

The nuclei $^{100}$Mo and $^{102}$Ru are important isotopes in the A~100 transitional regions. In this region, N = 58 (corresponding to $^{100}$Mo and $^{102}$Ru for Z=42, and Z=44, respectively) itself plays a central role. As illustrated in Fig. 3, the pairing gap values increase up to N~58, and after crossing from this neutron region, we see a decrease in the pairing gap values. also, nuclei with N ≤ 58 show vibrational structure at low energy ($R_{4/2}$ ≤ 2.4). Above N = 58, the structure tends to exhibit a rotational character. [74]. Therefore, abrupt changing in the nuclear structure was shown in Fig. 3 with the addition of a new parameter to show sudden changes, and this new parameter is the pairing gap that can be a new observable for CPS (shown in Fig. 1) in regions that are suitable for studying QPT.

In the theoretical calculations of the pairing gap quantities, according to Eq. (2), we need models that can explain the binding energy quantities, including $BE(Z, N), BE(Z, N-1), BE(Z, N+1)$. We can use some different models and formulas to obtain these quantities, but we need models that have a concept of the pairing gap within their structures and also use neutron and proton numbers as the parameters for the binding energy. Thus, we use some algebraic models that consider the structures of proton and neutron to be paired, and one of the best devices to describe the nuclear structure and shape phase transitions of nuclei is the IBM for even-even nuclei and IBFM for even-odd nuclei. In Ref. [80], the A = 100 region was represented with the IBM using a Hamiltonian that has constant parameters.

Both models should be used according to Eq. (2): IBM for even-even nuclei and IBFM for even-odd nuclei. The global part of the BE in the IBM ($BE^{gl}$) comes from that part of the Hamiltonian that does not affect the internal excitation energies. Can be written in terms of the total number of bosons, $N_B$, and its contribution to the BE reads as:

$$BE^{gl}(N_B) = E_0 + AN_B + \frac{B}{2}N_B(N_B - 1), \tag{5}$$

To avoid ambiguities, it is assumed in these expressions that $N_B$ always corresponds to the number of nucleon pairs, considered as particles, and is never considered as holes[78].

Also, by extracting BE from the eigenvalue of the Hamiltonian in IBFM [81], by considering $N_F = 1$ ($N_F$ represents the number of fermions) as:

$$N_F = 1, \quad BE^{gl}(N_B) = e_0 + e_1 N_B + e_2 N_B(N_B + 5) + e_3 + 4e_4 + e_5 N_B, \tag{6}$$

It should be noted that A, B, $E_0, e_0, e_1, e_2, e_3, e_4, e_5$ are constant for chains of isotopes (fixed Z) when the value of $N_B$ changes, except when crossing the mid-shell or passing between major shells, i.e. it provides a linear contribution[78]. We use experimental data about BE in Eqs. (5,6) and by the fitting method in Python, we obtained coefficients in Eqs. (5,6) approximately. Also, we show this coefficient in Table 1.

Table1. The coefficients of both models (IBM, IBFM) in a certain mid-shell range for Mo and Ru isotopic chains were obtained by a fitting method in Python.

| Mid-shell ranges | IBM(even-even) Coefficients(keV) | | | IBFM(odd-even) coefficients(keV) | | | | | |
|---|---|---|---|---|---|---|---|---|---|
| | $E_0$ | A | B | $e_0$ | $e_1$ | $e_2$ | $e_3$ | $e_4$ | $e_5$ |
| after mid-shell: $50 \leq N \leq 66$ | 8537.18 | 37.37 | -8.52 | 6994.66 | -221.64 | -4.01 | 6994.66 | -1357.5 | 275.21 |

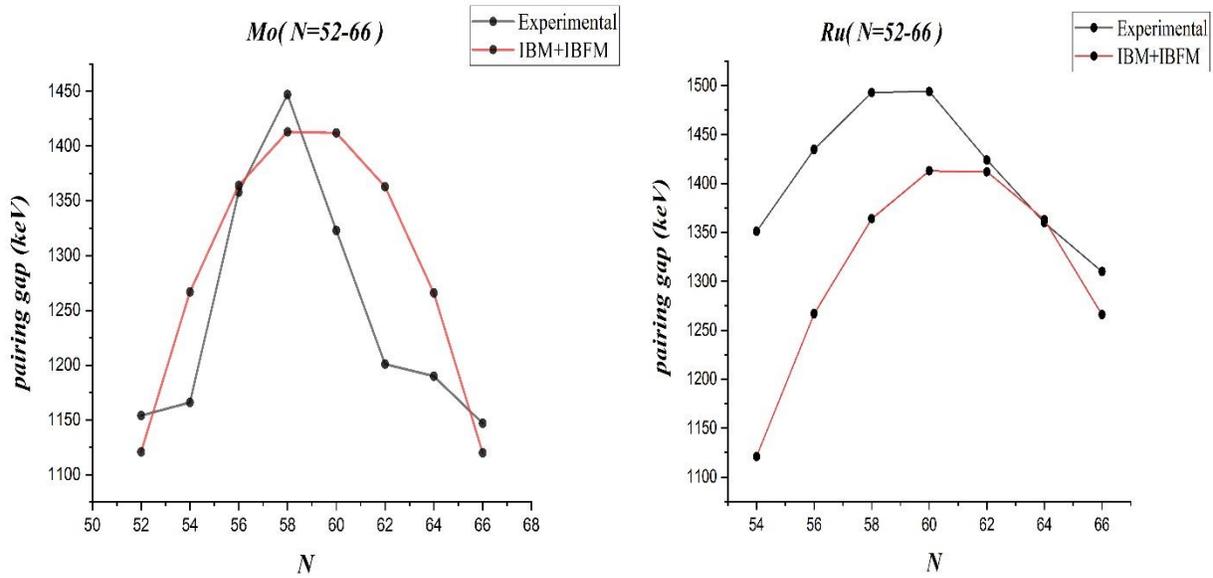

Figure 4. The variation of the experimental pairing gap with neutron number in the Mo and Ru isotopic chains is compared to the theoretical results for these elements.

The theoretical results in the IBM and IBFM framework (shown in Fig. 4) demonstrate a significant evolution of the pairing gap versus neutron number for Mo and Ru isotopic chains. This theoretical calculation also reveals abrupt changes near A≈100 and enhances our understanding of the phase transition in this important region.

Now, we will extend our idea about the pairing gap as a new observable to other isotopic chains located in different transitional regions, such as the Nd, Hf, Os, and Pt isotopic chains.

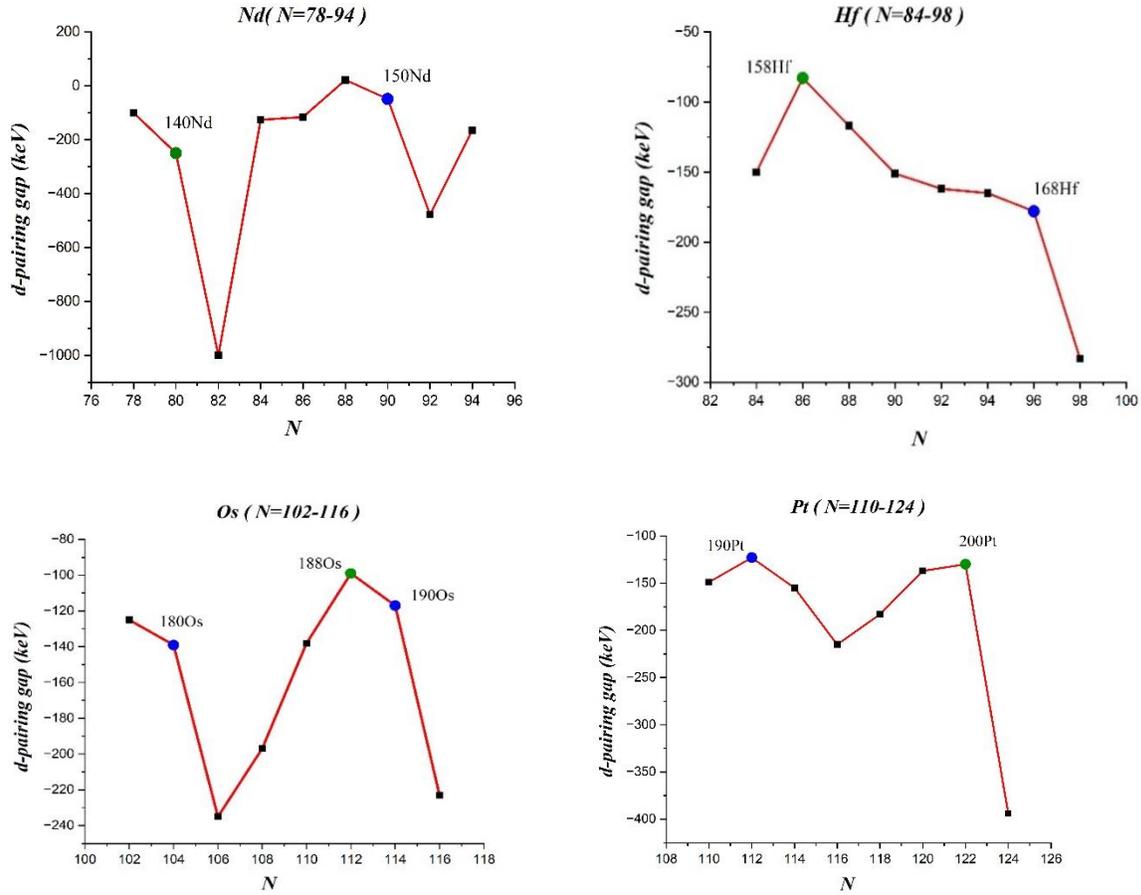

Figure 5. The variation of the experimental **d(pairing gap)** versus neutron number in Nd, Hf, Os, and Pt isotopic chains. Results of different studies that reported the possibility of critical points, shown with blue dots, and our suggestions for the possibility of critical points by using $R_{4_1^+/2_1^+}$ values, shown with green dots (taken from [73]).

Firstly, we focus on the reported nuclei as CPS in the A≈100. We have shown that we can find the critical points by variation of the d(pairing gap) versus neutron numbers. According to Figs. 1&5, the nucleus marked with blue dots represents the critical points that have been confirmed in different studies. The nucleus shown with green dots in Fig. 5 represents our suggested critical points, which have not been mentioned in previous studies. We have no claim that our candidates are as CPS, we just suggest some candidates based on two cases, in the first, we marked the blue and green dots in the same figures, not separately, because this method represents the similar variation and abrupt changing as CPS for the blue and green dots and in the second case, we investigated about the $R_{4/2}$ value and comparing this value with the range of E(5) and X(5) CPS, which confirms this nucleus as new candidates for critical points. We show in Table 2 that the nucleus, indicated by blue dots in Figs. 1&5 are candidates for critical points, which have been confirmed in different studies for nuclear structure.

Table 2. The references show the studies using other criteria for similar aims and different measures for similar results.

| Isotopic chain | Mo | Ru | Nd | Hf | Os | Pt |
| --- | --- | --- | --- | --- | --- | --- |
| Critical points for candidates | $^{100}$Mo | $^{102}$Ru | $^{150}$Nd | $^{168}$Hf | $^{180}$Os $^{190}$Os | $^{190}$Pt |
| references | [13,16] | [86,87] | [67,82-86,88,90] | [88-90] | [88,90,91] | [14] |

## 3. Summary and Conclusion

We utilized experimental data and algebraic models, including IBM and IBFM related to the pairing gap, to study CPS around nuclei that are near the Z≈40 and A≈100 region. The results of different studies which reported the possibility of QPT and Critical Points in the A≈100 region for Mo and Ru isotopic chains, also by comparing them in CPS, which can be seen in Fig. 1&5 (shown with blue dots), and in the following we extended our idea about the pairing gap as a new observable for CPS, to other isotopic chains that are located in the other transitional regions such as Nd, Hf, Os and Pt isotopic chains. Experimental investigations about the critical points by using the evolution of the d(pairing gap) confirm the results of [references in Table 2]. Also, our theoretical calculation corresponds to the results of the Mo and Ru isotopic chains (shown in Fig. 4), approximately. It should be noted that our results also describe other critical points that were not reported before this investigation. We investigated other CPS in different isotopic chains in Fig. 5 (shown with green dots) and according to the abrupt changing near transitional regions similar Fig. 3 and the $R_{4/2}$ value for E(5)~2.2 and X(5)~2.9, our suggestion candidates for E(5) and X(5) critical points including: $^{140}Nd$, $^{158}Hf$, $^{188}Os$, $^{200}P$t. Therefore near the magic numbers, it shows a stable and spherical state, while When we move away from the magic numbers, and approach the critical points, the pairing gap changes from its stable and symmetrical state to the deformed state, and the next symmetry state, so the pairing gap or d(pairing gap) parameter can be one of the observables of the critical points and QPT in the different nucleus. This experimental study on the evolution of the pairing gap highlights its significant role in the spectral properties of a nucleus and in understanding new nuclear structures. In theoretical investigations of the pairing gap, we can assess various observables, including $S_{2n}$ and neutron capture cross-section. So, the variation of these macroscopic signatures represents the evolution of the pairing gap, which is one of the microscopic phenomena. Therefore, by investigating both studies, we can show this important relation between the macroscopic and microscopic signatures.

## Acknowledgment

This work is supported by the Research Council of the University of Tabriz.

## Author contributions

H. Emami and H. Sabri performed the initial calculations, analyzed and interpreted the results, and wrote the main manuscript text. All authors commented on and reviewed the manuscript.

## Competing interests

The authors declare no competing interests.

## Data Availability Statement

The datasets used and analyzed during the current study are available from the corresponding author at the reasonable request.